\documentclass[prl,aps,twocolumn,showpacs,preprintnumbers,amsmath,amssymb,
superscriptaddress,notitlepage]{revtex4}

\usepackage{graphicx}
\usepackage{dcolumn}
\usepackage{bm}
scaled\magstep1

\let\a=\alpha \let\b=\beta    \let\d=\delta \let\e=\varepsilon
  \let\h=\eta   \let\th=\theta \let\k=\kappa \let\l=\lambda
\let\m=\mu    \let\n=\nu         \let\p=\pi    \let\r=\rho
\let\s=\sigma \let\t=\tau    
   \let\o=\omega
\let\G=\Gamma

\def\EE{{\cal E}} \def\VV{{\cal V}}
 
\def\NN{{\cal N}} 
\def\RR{{\cal R}}\def\LL{{\cal L}}

   \def\pp{{\bf p}}
 \def\xx{{\bf x}} \def\yy{{\bf y}} \def\zz{{\bf z}}

\def\kk{{\bf k}}

\def\\{\hfill\break}
\let\==\equiv
\let\io=\infty
\let\0=\noindent

\let\dpr=\partial

\def\tende#1{\,\vtop{\ialign{##\crcr\rightarrowfill\crcr\noalign{\kern-1pt
    \nointerlineskip} \hskip3.pt${\scriptstyle #1}$\hskip3.pt\crcr}}\,}
\def\otto{\,{\kern-1.truept\leftarrow\kern-5.truept\to\kern-1.truept}\,}

\def\to{\rightarrow}
\def\la{\left\langle}
\def\ra{\right\rangle}
\def\qed{\hfill\raise1pt\hbox{\vrule height5pt width5pt depth0pt}}

\def\be{\begin{equation}}
\def\ee{\end{equation}}
\def\bea{\begin{eqnarray}}
\def\eea{\end{eqnarray}}
\def\nn{\nonumber}
\def\pref#1{(\ref{#1})}

\begin{document}

\title{Conductivity in the Heisenberg chain with next to nearest neighbor interaction}
\author{Vieri Mastropietro}%
\affiliation{%
University of Milan, Math. Dept. "F. Enriquez", Via Saldini 50, Milan, Italy}

\begin{abstract} We
consider a spin chain given by the $XXZ$ model with a weak
next to nearest neighbor perturbation which breaks its exact integrability.
We prove that such system has an ideal metallic behavior (infinite conductivity),
by rigorously establishing strict lower bounds on the
zero temperature Drude weight which are strictly positive. The proof is based on
Exact Renormalization Group methods allowing 
to prove the convergence of the expansions and to fully take into account the irrelevant terms, which play an essential role in ensuring the correct lattice symmetries.
We also prove that the Drude weight verifies the same parameter-free
relations  as
in the absence of the integrability breaking perturbation.
\end{abstract}
\pacs{05.60Gg, 05.10.Cc, 75.10.Jm} \maketitle

\section{Introduction}\label{sec1}

The conductivity properties of quantum spin chains has been the subject of an intense
research in recent times, see e.g.  \cite{Z1,Z,P1,NMA,RA,RA1,AG,FK,HH,B,JR,HS,SPA1,SPA2,P,K1}, using both numerical
and analytical methods; however several basic aspects remain still controversial and this makes
the interpretation of experiments \cite{T,T1} problematic.
A prominent role among systems modeling spin chains is played by the $XXZ$ model, which was solved long ago by
Bethe ansatz \cite{YY} and describes spins with nearest neighbor perturbation.
If $S^\a_x = \s^\a_x/2$ for $i=1,2,\ldots,L$ and $\a=1,2,3$,
$\s^\a_x$ being the Pauli matrices, the Hamiltonian of the XXZ chain is $H_0=$
\be  -\sum_{x=1}^{L-1} [J S^1_x S^1_{x+1}+J S^2_x
S^2_{x+1} +J_3  S^3_x S^3_{x+1}+h S^3_x]+U_L  \label{1}
\ee
and $U_L$ takes into account boundary conditions. We will choose $J=1$ for definiteness.
The above model can be rewritten as a many body system of interacting spinless fermions through the
Jordan-Wigner transformation (see below); when $J_3=0$ the system is equivalent to model of free fermions.

An important question to be understood is how much  
the solvability property of the XXZ chain influences the conductivity.
Indeed in classical dynamics transport properties are very much affected by integrability
or nearly integrability, see {\it e.g.} \cite{BLR,L}, and one could wonder 
if the same happens in the quantum case. We can therefore add to the $XXZ$ chain
a {\it next to nearest neighbor interaction} breaking exact solvability, that is we consider 
the Hamiltonian \be H=H_0+H_1\label{1a1}\ee introduced in \cite{HA0}, with
\be H_1 =-\l\sum_{x=1}^{L-1} [ S^1_x S^1_{x+2}+S^2_x
S^2_{x+2}+  S^3_x S^3_{x+2}]\label{1a}\ee
Using the Peierls substitution we can derive the expression of the {\it spin current}, given by
\be
j_x=
S^1_x S^2_{x+1}-S^2_x S^1_{x+1}+\l F_x
\ee
where $F_x$ is an expression quartic in the spin operators whose explicit form will be written in
the following section.  If $\r_{x}=S^3_{x}+{1\over 2}$  and $(j^0_x,j^1_x)=(\r_x,j_x)$ we define
\be K^{\m,\n}_{\b,\l}(p_0,p)=\int_0^\b dx_0 e^{-i p_0 x_0}<\hat j^\m_{x_0,p} \hat j^\n_{x_0,-p}>_{\b,T}
\ee
and $<O>_\b={{\rm Tr} e^{-\b H}O\over {\rm Tr} e^{-\b H}}$, $O_{x_0}=e^{H x_0} O e^{-H x_0}$ and $T$ denotes truncation.
We will use the notation $\lim_{\b\to\io}<O>_\b=<O>$. An important thermodynamic quantity is
the susceptibility defined as 
$\k_{\l}=\lim_{p\to 0}\lim_{p_0\to 0}<\hat\r_{\pp} \hat\r_{-\pp}>$ with $\pp=(p_0,p)$.

The framework for most transport studies is linear response theory, where the conductivities 
are given in terms of dynamic correlations. According to Kubo formula, the {\it spin conductivity}
at frequency $\o$ at zero temperature is given by 
\be
\s_{\l}(\o)=\lim_{\d\to 0}\lim_{p\to 0}\lim_{\b\to\io}{D_{\b,\l}(\pp)\over ip_0}|_{i p_0\to \o+i\d}\label{dr1}
\ee
where $\pp=(p_0,p)$ and  
\be
D_{\b,\l}(\pp)=[K^{11}_{\b,\l}(\pp)+<j^D>_\b]\label{dr2}\ee
where $j^D$ is the diamagnetic term (see below) and $D_\l=\lim_{p_0\to 0}\lim_{p\to 0}\lim_{\b\to\io}D_{\b,\l}(\pp)$ is the zero temperature {\it Drude weight}
\cite{K}. At zero temperature a non vanishing Drude weight signals an ideal metallic behavior with infinite conductivity; this is what happens at the free fermion point
 $J_3=\l=0$.  In the $XXZ$ chain ($J_3\not=0,\l=0$), corresponding to interacting fermions, 
the Drude weight can be computed by Bethe ansatz \cite{YY},\cite{SS},\cite{GS}
and a non vanishing result is also found
\be
D_0={\pi\over\bar\m}{\sin\bar\m\over 2\m (\pi-\bar\m)}\quad\quad  
\cos \bar\m=-J_3
\ee
implying infinite zero temperature conductivity;  this should  be compared with other interacting 1D Fermi systems, like the Hubbard model at half filling, in which even an arbitrarily weak interaction induces an insulating behavior \cite{LW}.
Remarkably the XXZ Drude weight is non vanishing even at positive temperature, see
\cite{Z1} ($h\not=0$) and \cite{P} ($h=0$); the conserved quantities in the integrable XXZ chain
imply dissipationless current at finite temperature.
 
Much less is known about the the conductivity in presence of an integrability breaking terms as \pref{1a}. 
There is no consensus even on the basic question if integrability breaking terms make the conductivity finite or not at {\it non zero} temperature;
some groups have results supporting a {\it finite} conductivity 
\cite{Z,Z1,RA,RA1,HH,JR} while others get evidence of an {\it infinite} conductivity
\cite{AG, P1, HS,FK,K1}. The reason of this ambiguity is the subtle interplay between dangerously
irrelevant terms and conserved quantities, making the results particularly sensitive to regularizations or approximations.

All the technical problems making the understanding of the finite temperature conductivity properties
of the non integrable chain \pref{1a1} so difficult and the conclusions so uncertain appear already at {\it zero} temperature; there 
exist indeed {\it no rigorous results} on the conductivity for the non integrable spin chain \pref{1a1} even at $T=0$. Numerical analysis using Exact Diagonalization or Montecarlo 
have severe limitations due to the finite system size. On the other hand 
the Bethe ansatz solution does not furnish a good starting point for a perturbative analysis.  
Usually the zero temperature Drude weight is computed replacing the chain model by a continuum
Quantum Field Theory model (the Thirring or the Luttinger model)
which can be mapped in a boson gaussian model (bosonization) in which the Drude weight
can be explicitly computed. The difference between the continuum and the original lattice model
is in terms which are {\it irrelevant} in the Renormalization Group sense, which take into account 
lattice effects like Umklapp and non linear bands. Such irrelevant terms are crucial ones at $T>0$ (neglecting them the conductivity is infinite and
it has been conjectured that their presence can 
render the conductivity finite) but also  at $T=0$ they are important and cannot be neglected;
their contribution to the Drude weight is of the same order than the value found in the gaussian model, so that in order to get a lower positive bound 
for $D_\l$ one has to exclude cancellations.
More in general, in conductivity problems 
one cannot trivially appeal to an universality principle 
and results are strongly sensitive to the irrelevant terms, that is
to the details of the model.
An example in which this is particularly transparent
is given 
by a recent theorem \cite{GMP} proving the universality of graphene conductivity in presence of interaction, in which the 
role if irrelevant terms is crucial.

Our main result is the following theorem
\vskip.3cm
{\bf Theorem.}
{\it There exists $\e<1$ such that, if $|J_3|,
|\l|\le \e$ the zero temperature Drude weight is non vanishing 
and analytic in $J_3,\l$; moreover
\be
D_\l=K {v_{s,\l}\over \pi}\quad \k_\l={K\over \pi v_{s,\l}}\label{aa1}
\ee
with $K=$
\be
1-{1\over \pi v_{s,\l}}[(J_3+2\l)(1-\cos 2 p_F)+\l(1-\cos 4 p_F)+F]\nn
\ee
and $v_s=\sin(p_F)+\tilde F$, $\sin p_F=h$ and 
$|F|\le C\e^2, |\tilde F|\le C\e$.
}
\vskip.3cm

The above result rigorously establish for the first time that
the Drude weight is finite for the
anisotropic $XXZ$ chain perturbed by a weak next to nearest neighbor interaction \pref{1a}, so that 
the system behave at zero temperature as an ideal metal (infinite conductivity).
It is based on the techniques introduced in \cite{BM1} and it extends \cite{BM} 
which was limited to a very special anisotropic spin interaction. Besides providing a strict lower bound 
on the Drude weight, the above theorem provides the validity of the relation (following from \pref{aa1})
\be
{D_\l \over \k_\l}=v_{s,\l}^2\label{aass}
\ee
which is known to be true at $\l=0$ from Bethe ansatz \cite{YY,SS,GS} as 
\be
\k_0={\bar\m\over 2\pi}{1\over (\pi-\bar\m)}\sin\bar\m\quad\quad v_{s,0}={\pi\over\bar \m}\sin\bar\m
\ee
Note that for $\l=h=0$ and small $J_3$
$K^{-1}=2(1-{\bar\m\over\pi})=$ $K^{-1}=1+{2J_3\over\pi}+ O(J_3^2)$ and $v_s=1+O(J_3)$
in agreement with our formulas. The validity of \pref{aa1} at zero temperature 
for the spin chain \pref{1a1} was conjectured in \cite{Ha}, but its validity was only checked at $\l=0$ when the exact solution is valid. 

The proof of the theorem is based on Exact Renormalization Group (ERG) methods, see {e.g.} \cite{GM}, which appears quite well suited for the problem;
contrary to the usual field theoretical RG, in which the irrelevant terms are simply neglected, in ERG no approximations are
done and all the irrelevant terms, which are crucial in conductivity properties, are fully taken into account.

\section{Lattice Ward Identities}

The spin chain Hamiltonian can be rewritten as a fermionic Hamiltonian by using the Jordan-Wigner transformation;
calling $S^+_x=S^1_x+i S^2_x$ and  $S^-_x=S^1_x-i S^2_x$ we can write
\be
S^-_x=e^{-i\pi\sum_{y=1}^{x-1}a^+_y a^-_y}a^-_x \quad  S^+_x=a^+_x e^{i\pi\sum_{y=1}^{x-1}a^+_y a^-_y}
\ee
and $S^3_x=a^+_x a^-_x-1/2$, where $a^\pm_x$ are fermionic creation and annihilation operators. 
One finds
\bea&& H_0 = -\sum_{x=1}^{L-1} {1\over 2} [ a^+_{x} a^-_{x+1}+
a_{x+1}^+
a^-_{x}] -h\sum_{x=1}^L ( a^+_x a^-_x-{1\over 2}) +\nn\\
&& -J_3 ( a^+_x a^-_x-{1\over 2})(
a^+_{x+1} a^-_{x+1}-{1\over 2})\; \eea
and
\bea &&H_1=-\l\sum_{x=1}^{L-1} [- a^+_x ( a^+_{x+1}
a^-_{x+1}-{1\over 2})a^-_{x+2}-\\
&
&a^+_{x+2} ( a^+_{x+1} a^-_{x+1}-{1\over 2})a^-_{x}+(
a^+_{x}
a^-_{x}-{1\over 2})(a^+_{x+2} a^-_{x+2}-{1\over 2})] \nn\; \eea
and $U_L$ can be chosen so that $a^\pm_x$ verify periodic boundary conditions.
The above representation makes straightforward the computation of the spin currents
which are e.m. currents
in the fermionic representation.
Using the Peierls substitution, the coupling with a classical e.m. field is described by the Hamiltonian
$H+H(A)$
with
\bea
&&H(A)=-\sum_{x=1}^{L-1} {1\over 2}  [a^+_{x} U_{x,x+1}
a^-_{x+1}+a^+_{x+1} U_{x+1,x}
a^-_{x}]+\nn\\
&&\l\sum_{x=1}^{L-1} [a^+_x ( a^+_{x+1}
a^-_{x+1}-{1\over 2}) U_{x,x+2}a^-_{x+2}+\\
&&a^+_{x+2} ( a^+_{x+1} a^-_{x+1}-{1\over 2})U_{x+2,x
}a^-_{x}]\nn
\eea
and $U_{x,y}=e^{i\int _x^y ds A_1(s) }-1$. The {\it paramagnetic current} is, if $\h_{1,p}={1-e^{-ip}\over i p} $ and $\h_{2,p}={1-e^{-2 ip}\over 2 i p} $
\bea
&&
\hat j_{p}=-{d H(A)\over d A_p}|_{0}=-{1\over L}\sum_k \h_{1,p} 
({e^{-i(k+p)}-e^{i k}\over 2i}
)\hat a^+_{k} \hat a^-_{k+p}\nn\\
&&-{\l\over L}\sum_k \h_{2,p}
({e^{-2i(k+p)}-e^{2i k}\over 2i} )\hat a^+_{k}\hat a^-_{k+p}+\\
&&-{\l\over L^3}\sum_{k_1,k_2,k_3}
\h_{2,p} w(\underline k,p)
\hat a^+_{k_1}\hat a^+_{k_2}\hat a^-_{k_3}\hat a^-_{k_1+k_2-k_3+p}\nn
\eea
with $w(\underline k,p)={1\over 2i}(
e^{-2i k_1-i k_2+i k_3-
2i p}-e^{+2i k_1+i k_2-i k_3})$.

Similarly the {\it diamagnetic current} is defined as
\be
j^D={\partial^2 H(A)\over \partial A_p \partial A_{-p}}|_{0}\ee
Note that the current is not anymore quadratic as in the $\l=0$ case.
 
The vertex and the two point correlation are connected by the following Ward Identities
 \bea&&-i p_0
<\hat\r_\pp\hat a^-_{\kk} \hat a^+_{\kk+\pp}>_{\b,T}+p
<\hat j_\pp\hat a^-_{\kk}\hat a^+_{\kk+\pp}>_{\b,T}
=\nn\\ &&[\la\hat a^+_{\kk}\hat
a^-_{\kk}\ra_{\b,T}-\la\hat a^+_{\kk+\pp}\hat
a^-_{\kk+\pp}\ra_{\b,T}]\label{ao1}\eea
Similarly the current-current and density-density correlations obey to
\bea
&&-i p_0 \hat K^{0,0}_{\b,\l}(\pp)+p \hat K^{10}_{\b,\l}(\pp)=0\nn\\
&&-i p_0 \hat K^{0,1}_{\b,\l}(\pp)+p (\hat K^{1,1}_{\b,\l}(\pp)+ <j^D>_\b)=0\label{ao2}
\eea
Let us consider the $\b\to\io$ limit.
If $\hat K_{00}(p_0,0)$ and $\hat D(\pp)$ were continuous in $\pp=0$, \pref{ao2}
would imply that both $\k$ and $D$ are vanishing. In the case we are
considering, we will see in the next section that $\hat K_{00}(p_0,0)$
and $\hat D(\pp)$ are bounded but not continuous in $\pp=0$, and this fact implies
the following identities
\be \hat K^{00}_\l(p_0,0)=0,\quad  [\hat K^{1,1}_{\l}(\pp)+<j^D>]_{p_0=0}=0\label{ffvv}
\ee
The simplest derivation of the above Ward Identities is through to the Grassmann integral representation
for the correlations.
We introduce the generating function
\be e^{W(A,\phi)}=\int P(d\psi)
e^{\VV(\psi)+(\psi,\phi)+B(A,\psi)}\label{por} \ee
where, if $\kk=(k_0,\vec k)$ with $k_0$ the Matsubara frequency,
$P(d\psi)$ is the fermionic gaussian integration with
propagator
\be \hat g(\kk)={1\over -i k_0+\cos k-\m}\label{sas} \ee
with $\m=h+\l+J_3$ and
\bea  &&\VV=\int_0^\b dx_0\sum_{x} [\l (\psi^+_{\xx}\psi^-_{\xx+2 {\bf e_2}}+\psi^+_{\xx+2{\bf e_2}}
\psi^-_{\xx})+\label{ff}\\
&&
J_3\psi^+_{\xx}\psi^-_{\xx}\psi^+_{\xx+{\bf e}_1}
\psi^-_{\xx+{\bf e}_1}-\l\psi^+_{\xx}\psi^+_{\xx+{\bf e}_1}\psi^-_{x+{\bf e}_1}
\psi^-_{\xx+2{\bf e}_1}-\nn\\
&&\l\psi^+_{\xx+2{\bf e}_1}\psi^+_{\xx+{\bf
e}_1}\psi^-_{\xx+{\bf e}_1}
\psi^-_{\xx}+\l\psi^+_{\xx}\psi^-_{\xx}\psi^+_{\xx+2{\bf
e}_1}\psi^+_{\xx+2{\bf e}_1}]\nn\eea
Finally the source term is
given by
\bea && B(A,\psi)=\int_0^\b dx_0\sum_{
x}\Big[\psi^+_{\xx}\psi^-_\xx A_0(\xx)+\nn\\
&& [\psi^+_{\xx}
(e^{i\int_x^{x+1}A_1(s)}-1)\psi^-_{\xx+{\bf e}_1}+ \psi_{\xx+{\bf
e}_1}^+ (e^{i\int_{x+1}^{x}A_1(s)}-1)\psi^-_{\xx}]+\nn\\
&&\l \int_0^\b dx_0\sum_{x}
\l\psi^+_{\xx}(e^{i\int_{x+2}^{x}A_1(s)}-1)\psi^-_{\xx+2{\bf
e}_1}\nn\\
&&+\l \psi^+_{\xx+2{\bf
e}_1}(e^{-i\int_{x+2}^{x}A_1(s)}-1)\psi^-_{\xx}+\\ &&
\l(e^{i\int_x^{x+2}A_1(s)}-1)\psi^+_{\xx}\psi^+_{x+{\bf
e}_1}\psi^-_{x+{\bf e}_1} \psi^-_{x+2{\bf
e}_1}+\nn\\
&&\l (e^{-i\int_x^{x+2}A_1(s)}-1)\psi^+_{\xx+2{\bf
e}_1}\psi^+_{x+{\bf e}_1}\psi^-_{x+{\bf e}_1} \psi^-_{x}\Big]
\label{1.9}\eea
and \be (\psi,\phi)=\int_0^\b dx_0\sum_{x}
[\psi^+_\xx\phi_\xx^- +\phi^+_\xx\psi^-_\xx]\ee
The correlations are easily written in terms of derivatives of the generating function; 
in particular
\bea
&&<\hat j_p\hat a^-_{\kk} \hat a^+_{\kk+\pp}>_{\b,T}={\partial^3 W(A,\phi)\over \partial A^1_{\pp}
\partial\phi^+_{\kk} \partial\phi^-_{\kk+\pp} }|_0\nn\\
&&\hat K^{0,0}_{\b,\l}(\pp)={\partial^2 W(A,\phi)\over \partial A_{0,\pp}\partial A_{0,-\pp}}|_0\\
&&\hat K^{1,1}_{\b,\l}(\pp)+<j^d>_\b={\partial^2 W(A,\phi)\over \partial A_{1,\pp}\partial A_{1,-\pp}}|_0\nn
\eea
and so on. Performing the phase transformation \be \psi^{\pm}_\xx\to e^{\pm i 
\a_\xx}\psi^{\pm}_\xx\ee in(\ref{por}), we find
\be W(A+\dpr\a,\phi e^{i\a})=W(A,\phi)\;, \label{1.10}\ee
Therefore
by performing derivatives with respect to $\a$ and to the external fields $A, \phi$ the Ward Identities
\pref{ao1},\pref{ao2} follow.

\section{Exact Renormalization Group analysis}

The perturbation theory for the correlation functions is
(apparently) affected by infrared divergences, related to the divergence of the
free propagator Eq.(\ref{sas}) at $\cos k=\m$. As the interaction modifies in general the location of the singularity
it is convenient to write $\m=\cos p_F+\n$, where $\n$ is a counterterm fixed so that the singularity of the two point function  
is at $k=p_F$. We consider the following equivalent generating functional
\be e^{\tilde W(A,\phi)}=\int P(d\psi)
e^{\VV(\psi)+(\psi,\phi)+\bar B(A,\psi)}\label{por1} \ee
where $P(d\psi)$ has now propagator
\be g(\kk)={1\over -i k_0+{v_s\over v_F}(\cos k-\cos p_F)}\label{sas1} \ee
where $v_s=\sin p_F(1+\d)$, $v_F=\sin p_F$ and
in $\VV$ there are two new quadratic term proportional to $\n,\d$; moreover 
$\bar B(A,\psi)$ is a source term given by $
\bar B(A,\psi)=\int d\xx [A^0 \r_\xx+A^1 j_\xx]$ (we just keep the linear part for the computation of the current-current correlation).
The source term $\hat A_\pp$ is assumed with a compact support and
we will choose $\n,\d$ as function of $J_3$ and $\l$ so that the Fermi point of the interacting theory is just $p_F$ and the velocity $v_s$. 

The functional integral \pref{por} is perfomed in a multiscale fashion using the following two basic properties of Grassmann integration. The first is the {\it addition property}, which says that
\be
\int P(d\psi) F(\psi)=\int P(d\psi^{(1)}) \int P(d\psi^{(2)}) 
F(\psi^{(1)}+\psi^{(2)})\label{add1}\ee
where $P(d\psi) ,P(d\psi^{(1)}),P(d\psi^{(2)})$ are Grassmann integrations 
with propagators $g, g^{(1)}, g^{(2)}$ with $g=g^{(1)}+g^{(2)}$ and $F(\psi)$
is an analytic function in $\psi$. The other is the {\it invariance of the exponential}, which says that
\be
\int P(d\psi)e^{V(\psi+\phi)}=e^{V'(\phi)}\label{add2}
\ee
with
\be
V'(\phi)=\sum_{n=1}^\io{1\over n!}\EE_\psi^T(V;n)\label{add3}
\ee
and $\EE^T$ are the fermionic truncated expectation. 
Using \pref{add1} we can decompose 
$\psi$ as a sum of independent Grassmann
fields $\psi^{(h)}$
living on momentum scales $|\kk-\pp_F^\pm|\simeq 2^{h}$,  $\pp_F^\pm=(0,\pm p_F)$ with
$h\le 0$ a scale label and $\cos p_F=\m$. 
After the integration of the fields with scales $\ge h$ we rewrite
Eq.\pref{por1} (setting, for simplicity, $\phi^\pm=0$) as
$e^{\tilde W(A,\phi)}=$
\be
\NN_h\int \prod_{\o=\pm}P_{Z_h}(d\psi_\o^{(\le h)})
e^{\VV^{(h)}(\psi^{(\le h)})+B^{(h)}(A,\psi^{(\le h)}) }\;,\label{ka}\ee
where $\psi^\pm_{\kk',\o}$, $\o=\pm 1$ is the quasi-particle field at the Fermi point
$\pp_F^\o$ (with quasi-momentum $\kk'$ relative to the Fermi point $\pp_F^\o$)
and $P_{\le h}(\psi_{\o})$ is a fermionic
gaussian integration with propagator
\be g_{\o}^{(\leq h)}(\kk')=\frac{1}{Z_h}
{\chi_h(\kk')\over -i k_0+\o v_s \sin k' +\cos p_F (\cos k-1)}\label{gfg}
\ee
where $\chi_h(\kk')$ is a cut-off function supported in $|\kk'|\le 2^h$
and $Z_h$ 
is the effective wave function renormalization. The {\it effective potential} $\VV^{(\le h)}(\psi^{(\le h)})$ is a
sum of integrals of monomials in $\psi^{(\le h)}$ of order $n$
multiplied by kernels
$W^{(h)}_{n,0}(\xx_1,\ldots,\xx_n)$; similarly the effective source is
$B^{(h)}(A,\psi^{(\le h)})$ is sum of integrals of monomials 
with $n$ $\psi$-fields and  $m$ A-fields multiplied by kernels
$W^{(h)}_{n,0}$. 

The scaling dimension is given  by, if $m_4$ is the number of quartic interactions, $n_{2,A}$ and $n_{4,A}$
are the number of source terms with two and four fermionic lines and $n_e$ the number of fermionic external lines
\bea
&&-2(m_4+n_{2,A}+n_{4,A}-1)+\\
&&(2 m_4+2 n_{4,A}+n_{2,A}-{n_e\over 2})=2-n_{2,A}-{n_e\over 2}\nn
\eea
Therefore, the {\it marginal} terms
in the RG sense are those with $n_e=4$, $n_A=0$, 
$n_e=2$, $n_A=1$, and the relevant are the ones with 
$n_e=2$, $n_A=0$.
All the other terms are {\it irrelevant}, in particular the terms
with six or more fermionic fields, 
corresponding to the effective
multi-particle scattering terms, or the source terms with 
more than  two fermionic fields. 
The integration of the scale $h$ is done, using the addition property \pref{add1}, writing
\bea && \NN_{h}\int \prod_{\o=\pm}P_{Z_{h-1}}(d\psi_\o^{(\le h-1)})\label{00}\\
&&\int \prod_{\o=\pm}P_{Z_{h-1}}
(d\psi_\o^{(h)})
e^{\LL\VV^{(h)}+\RR\VV^{(h)}+\LL B^{(h)}+\RR  B^{(h)}  }\nn\eea
with $\RR=1-\LL$ and $\LL\VV^{(h)}$ contains the marginal and relevant part of the effective interaction
\be
\LL\VV^{(h)}=\n_h 2^h F_h^\n+\d_h F_h^\d+\l_h F_h^\l
\ee
where 
\bea
&&
F_h^\n=\int_{-{\b\over 2}}^{\b\over 2} dx_0 \sum_{\o=\pm}\sum_x \psi^+_{\xx,\o} \psi^-_{\xx,\o}\nn\\
&&
F_h^\d=\int_{-{\b\over 2}}^{\b\over 2}dx_0\sum_{\o=\pm}\sum_x \psi^+_{\xx,\o} \o v_s \partial_x \psi^-_{\xx,\o}\nn\\
&&F_h^\l=\int_{-{\b\over 2}}^{\b\over 2}dx_0\sum_x \psi^+_{\xx,+} \psi^-_{\xx,+}
\psi^+_{\xx,-} \psi^-_{\xx,-}
\eea
while in $\RR\VV^{(h)}$ are all the {\it irrelevant terms}. In the same way  
\be B^{(h)}(A,\psi)= \sum_{\m=0}^1 Z_{\m,h}\int
\frac{d\pp}{(2\p)^3}A_{\m}(\pp) J_{\m}(\pp)\label{add22}
\ee
with $J_0=\sum_{\o=\pm 1} \psi^+_{\o,\xx}\psi^-_{\o,\xx}$, $J_1=\sum_{\o=\pm 1} \o \psi^+_{\o,\xx}\psi^-_{\o,\xx}$.
Starting from \pref{00}, we can integrate the field $\psi^{(h)}$ using \pref{add1}
and \pref{add2} and the procedure can be iterated. 

Each single scale propagator $g^{(h)}(\xx)$, given by \pref{gfg} with $\chi_h(\kk')$
replaced by $f_h(\kk')$,  a smooth function non vanishin only in $2^{h-1}\le |\kk'|\le 2^{h+1}$,
verifies the following bound, for any
integer $N$
\be
|g^{(h)}(\xx)|\le 2^h {C_N\over 1+[2^h|\xx|]^N}\label{ss}
\ee
implying
\be
|g^{(h)}(\xx)|_{L_\io}\le C 2^h\quad\quad |g^{(h)}(\xx)|_{L_1}\le C 2^{-h}
\label{bon}\ee
The outcome of the above construction is that the kernels $W^{(h)}_{n,m}$ 
are expressed as series in the running coupling constants $\vec v_k=(\n_k,\l_k)$, $k\ge h$. The fermionic expectations in \pref{add3}
are expressed in terms of determinants, and expanding them one 
obtains a Feynman graph representation
for the kernels  $W^{(h)}_{n,m}$. The Feynman graphs are finite
uniformly in $h$ (this would be not true in the {\it non renormalized expansion}
$\LL=0$); however their number
grows as $O(l!^2)$, if $l$ is the order, so that in this way can only prove that the l-th order is bounded by $C^l l! [max_{k\ge h} |\vec v_k|]^l$
(a result usually called $n!$ bound ),  a result which is not sufficient to establish the convergence of the series. In order to improve such bound one has to notice that
the fermionic expectation are given by
\be
\EE_h(\psi^-_{\yy_1}...\psi^-_{\yy_n}\psi^+_{\xx_1}...\psi^+_{\xx_n})={\rm det} (g^{(h)}(\xx_i-\yy_j))\label{ff}
\ee
and the fermionic propagator can be written in the form
\be
g^{(h)}(\xx-\yy)=\int_{-{\b\over 2}}^{\b\over 2} dz_0 \sum_z \, A_h^*(\xx-\zz)\cdot B_h(\yy-\zz)\label{xxx1},
\ee
with
\be ||A_h||^2=\int_{-{\b\over 2}}^{\b\over 2} dz_0 |A_h(\zz)|^2\le C 2^{-2h}\;,\quad\quad
||B_h||^2\le C 2^{4h}\ee
for a suitable constant $C$.
According to the {\it Gram inequality}, if a $n\times n$ matrix $H$ has the form 
$H_{ij}=({\bf f}_i,{\bf g}_j)$ then 
\be
|{\rm Det} H|\le ||{\bf f}||^n ||{\bf g}||^n
\ee
Therefore, while expanding the determinant in \pref{ff} 
and bounding each term a bound
$C^n n! 2^{n h}$ is found, using the Gram inequality one gets an estimate without factorials, namely $C^n  2^{n h}$. In fact in the kernels not the {\it simple}
expectations $\EE$ but the truncated ones $\EE^T$ appear; however
one can use the Battle-Brydges-Federbush formula allowing to write the truncated expectations as sum over chains of propagators (ensuring the connection)
times determinants which can be bounded by the Gram inequality.
Therefore, see Theorem 3.12 of
\cite{BM1} for the proof, by using the above ideas it is proved that 
the kernels $W^{(h)}_{n,m}$ are analytic in the running coupling constants (with a small but uniform in $h$ radius of convergence)  and 
satisfy
the bounds 
\be \int|W^{(h)}_{n,m}|\le C^{n+m}
|\l_h|^m2^{(2-n/2-m)h}\;, \label{1.15}\ee
Of course all the consistency of the method rely on the fact that the running coupling constants remain in the analyticity radius. It has indeed proved in \cite{BM2} that, by suitably choosing $\n,\d$, see \cite{BM2}
\be
\l_h\to \l_{-\io}(\l,J_3)\quad\quad \d_h\to 0, \n_h\to 0
\ee
where $\l_{-\io}(\l)$ are {\it analytic} functions of the coupling; in other words, there is a {\it line of fiixed points}.
This is consequence of a property, called  vanishing of the Beta function, which has been proved in \cite{BM2}
by implementing Ward Identities at each Renormalization Group iteration.

Another crucial property is that 
the wave function renormalization $Z_h$ appearing in \pref{00} and the vertex renormalizations $Z_{0,h}, Z_{1,h}$
appearing in \pref{add22} have an anomalous behavior consisting in a power law divergence $Z_{h}=O(2^{\h h})$ 
and $Z_{i, h}=O(2^{\h_i h})$, with $\h,\h_i$ positive and $O(\l_0^2)$; remarkably such exponents are {\it equal} $\h=\h_0=\h_1$, that is
\be
{Z_{0,h}\over Z_h}\to 1+A(\l,J_3)\quad \quad {\tilde Z_{1,h}\over Z_h}\to 1+B(\l,J_3)\label{zaza}
\ee
and $A, B=O(\l_0)$, and $\h=a\l_0^2+O(\l_0^3)$. The identity of such exponents is due to emerging relativistic symmetries (see below) which are broken by the irrelevant terms; this is reflected by the fact that
$A\not=B$ as one can verify by an explicit computation. Such emerging relativistic symmetries emerges from the following
decomposition of the fermionic propagator 
\be
g^{(h)}_\o(\xx)=g^{(h)}_{\o,R}(\xx)+r^{(h)}_{\o}(\xx)\label{dec}
\ee
with
\be
g^{(h)}_{\o,R}(\xx)={1\over \b L}\sum_{\kk} {f_h(\kk)\over -i k_0+\o k v_s}
\ee
and 
\be
|r^{(h)}_{\o}(\xx)| \le C { 2^{2h}\over 1+[2^h|\xx|]^N}
\ee
for any $N$, that is with an extra $2^h$ with respect to the bound for the dominant part. The above decomposition says that the single scale propagator is closer and closer to
the one of a massless Dirac fermions, that is $g^{(h)}_{\o,R}(\xx)$, plus a correction
$r^{(h)}_{\o}(\xx)$ taking into account of the non linear bands. Of course even if the relative size
of the two terms in \pref{dec} is asymptotically vanishing,
the contribution of the $r^{(h)}_{\o}(\xx)$ terms to the thermodynamical constants is not negligible. The 
current current correlation is naturally decomposed in two terms;
one, denoted by $K^{(a)11}_\l(\xx)$, which takes contributions from the non irrelevant part of the effective potential 
$\LL\VV^{(k)}$ and from the "relativistic" part of the propagator
$g^{(h)}_{\o,R}(\xx)$ plus a rest 
$K^{(b)11}_\l(\xx)$ depending from the irrelevant terms
\be
K^{11}_\l(\xx)=K^{(a)11}_\l(\xx)+K^{(b)11}_\l(\xx)\label{sac}
\ee
where 
\be
K^{(b)11}_\l(\xx)=\sum_{h=-\io}^0  S_R^{(h)}(\xx)\label{sacc}
\ee
and
\be
|S^{(h)}_R(\xx)|\le 2^{2h}2^{\th h}{1\over 1+[2^h |\xx|]^N}\label{sacc1}
\ee
where the extra $2^{\th h}$, $\th=1/2$,  with respect to the dimensional bound is due to the fact that it has contribution from the irrelevant terms. From \pref{sacc1} 
and \pref{sacc} we obtain the following bound
\be
|K^{(b)11}_\l(\xx)|\le {C\over 1+|\xx|^{2+\th}}
\ee
from which we deduce that the Fourier transform $\hat K^{(b)11}_\l(\pp)$
is continuous in $\pp$ and $O(1)$. There is therefore no justification in neglecting
the irrelevant terms, as they give an $O(1)$ contribution to the Drude weight.

On the other hand $K^{(a)11}_\l(\xx)=$
\bea
&&\sum_{h=-\io}^0 \sum_{\o=\pm }\{[{Z_{1,h}\over Z_h}]^2 g^{(h)}_{\o, R}(\xx) 
g^{(h)}_{\o,R}(-\xx)(1+\G_h(\xx))\label{aassaa}\\
&&
+\cos (2p_F x)
[{Z_{2,h}\over Z_h}]^2 g^{(h)}_{\o, R}(\xx) 
g^{(h)}_{-\o,R}(-\xx)(1+\tilde\G_h(\xx))\}
\nn
\eea
where $|\partial^n \G_h(\xx)|,|\partial^n \tilde\G_h(\xx))|\le C|\l_0|2^{-n h} $, 
and $Z_{2,h}\sim 2^{\tilde\h h}$ with $\tilde\h=b\l_0+O(\l_0^2)$.
We are interested to the Fourier transform $\hat K^{(a)11}_\l(\pp)$; the contribution from the oscillating part
of \pref{aassaa} is surely bounded close to $\pp=0$ but it is not obvious at all
that Fourier transform of the first term (the non oscillating one) is bounded. It behaves
for large distances, from  \pref{zaza}, as $O(|\xx|^{-2})$ and logarithmic divergences could be present. In conclusion, the exact Renormalization Group analysis
provides non perturbative bounds for the current-current correlations in the coordinate space; we cannot deduce by such bounds even that the Drude weight is finite, as
logarithmic divergences could be present. Even if we could prove that 
$\hat K^{(a)11}(\pp)$ is bounded in the limit this would not allow us to conclude
anything on the Drude weight for the presence of $\hat K^{(b)11}_\l(\pp)$ which is $O(1)$.

\section{Emerging chiral symmetries}

The dimensional bounds obtained from the Renormalization Group analysis are not sufficient
for the Drude weight and one needs
to exploit both the lattice and the emerging chiral symmetries of the theory.
In order to do that, and remembering \pref{dec},
we introduce a model expressed of massless Dirac fermions in $d=1+1$ with light velocity $v_s$. The generating functional is given by 
\be
e^{W_{rel}(B,\phi)}=\int P(d\psi)e^{\VV(\sqrt{Z}\psi)+(\psi,\phi)+{\cal B}(B,\psi)}\label{aaaa}
\ee
where $P(d\psi)$ is the fermionic integration with propagator 
 \be g_\o(\kk)= {1\over Z}
{\chi_N(\kk)\over -i k_0+ \o v_s k}\ee with $\chi_N(\kk)$
a cut-off function selecting moments less than $2^N$, \bea &&V=\tilde\l_\io\int d\xx \int d\yy v(\xx-\yy)\psi^+_{+,\xx}
\psi^-_{+,\xx}\psi^+_{-,\yy}
\psi^-_{-,\yy}\nn\\
&&{\cal B}(B,\psi)=\sum_{\m=1}^1 B_\m \tilde Z^{(\m)} J_\m\eea
with $v(\xx)$ a short ranged interaction $|\hat v(\pp)|\le C e^{-\k |\pp|}$,
$J^0_\xx=
(\psi^+_{\xx,+}
\psi^-_{\xx,+}+\psi^+_{\xx,-}
\psi^-_{\xx,-})$ 
and $J^1_\xx=(\psi^+_{\xx,+}
\psi^-_{\xx,+}-\psi^+_{\xx,-}
\psi^-_{\xx,-})$. Note that in this case $\xx$ is a continuum variable both in space and time while in the previous case
the the spatial component was discrete. The Schwinger functions are given by derivatives of the generating function 
\bea &&\hat G^{2,1}_{\m,\o}(\kk,\kk+\pp)={\partial^3 \tilde W_{rel}\over \partial B_{\m,\pp} \partial\phi^+_{\o,\kk}
\partial\phi^-_{\o,\kk+\pp}}\quad i=1,2\nn\\
&&\hat G^{0,2}_{\m,\n}(\pp)={\partial^2 \tilde W_{rel}\over \partial B_{\m,\pp}\partial  \partial B_{\n,-\pp}
}\quad \m,\n=1,2
\eea
We can analyze the above functional integral using a Renormalization Group analysis. With respect to the previous case, in which the lattice
furnishes an {\it ultraviolet} cut-off and the problem is an {\it infrared} one (that is the zero temperature and infinite volume limit),
in the model \pref{aaaa} there is both an {\it ultraviolet} and {\it infrared} problem. We write then $\psi=\sum_{-\io}^N \psi^{(h)}$
where $\psi^{(h)}$ lives on momentum scale 
$|\kk|\simeq 2^{h}$. Note the crucial difference with respect to the chain model described above; in the present case 
there are positive scales (the momentum $\kk$ is unbounded in the $N\to\io$) while in the presence of the lattice the scale are $h\le 1$ as
$k \in [-\pi, \pi]$ for the presence of the lattice (there is only an innocuous ultraviolet problem for the unboundedness of $k_0$). The integration of the positive ultraviolet scales has been analyzed in \cite{M}, and it has proved that 
after the integration of the fields $\psi^{(N)},\psi^{(N-1)},..,\psi^{(1)}$ one gets an effective potential with kernels {\it uniformly bounded} as $N\to\io$.
A crucial role in establishing this result relies on the {\it non-locality} of the interaction, which eliminates the possible ultraviolet divergences 
present in the case of local $\d$-like interactions. Once that the ultraviolet scales are integrated out, the integration of the negative infrared scales
is done as described above for the spin chain, with some obvious modification due to symmetry; 
we call the corresponding effective couplings $\tilde\l_h, \tilde Z_{0,h}, \tilde Z_{1,h}$ 
while due to symmetry $\d_h=\n_h=0$. 
Again the beta function is vanishing and the effective coupling tends to a line of fixed points 
$\tilde\l_h\to_{h\to-\io}\tilde\l_{-\io}$, with $\tilde\l_{-\io}$ an analytic function of $\tilde\l_{\io}$.

The model \pref{aaaa} can be considered the continuum limit of the chain model.
It verifies more symmetries than the chain model; for instance it is symmetric with respect to space-time inversion and it invariant under the {\it chiral} transformation
$\psi^\pm_{\o,\xx}\to e^{\pm i\a_\o}\psi^\pm_{\o,\xx}$. The advantage of our ERG method is that the relation between the two models can be understood {\it quantitatively} in the following precise sense: 
it is possible to choose $\tilde\l_\io$, $Z,\tilde Z^{(0)},\tilde Z^{(1)}$ so that, for small $\pp,\kk'$, $\o=\pm$
\bea 
&&<a^-_{\kk'+\pp_F^\o}a^+_{\kk'+\pp_F^\o}>=G^{2,0}_{\o}(\kk')(1+O(\kk'))\\
&&<\hat\r_\pp\hat a^-_{\kk'+\pp_F^\o} \hat a^+_{\kk'+\pp+\pp_F^\o}>_{T}=
G^{2,1}_{0,\o}(\kk',\kk'+\pp)(1+R_1)\label{cczs}\nn\\
&&<\hat j_\pp\hat a^-_{\kk'+\pp_F^\o} \hat a^+_{\kk'+\pp+\pp_F^\o}>_{T}=
G^{2,1}_{1,\o}(\kk',\kk'+\pp))(1+R_2)\nn\eea
with $R_1, R_2=O(|\kk'|,\pp)$ and
\be \hat K^{\m,\n}_{\l,\b}(\pp) = \hat
G^{0,2}_{\m,\n}(\pp)+\hat A_{\m,\n}(\pp)\label{ffbb}
\ee
with $A_{\m,\n}(\xx)\le C|\xx|^{-2-\th}$, hence $\hat A_{\m,\n}(\pp)$ is continuous in $\pp$ at $\pp=0$.
This can be proved choosing $\tilde\l_{\io}, Z,\tilde Z^{(0)},\tilde Z^{(1)}$, by the implicit function theorem, 
so that the differences between $\tilde\l_h,\tilde Z_h,\tilde Z_{0,h}, \tilde Z_{1,h}$
and $\l_h, Z_h,Z_{0,h}, Z_{1,h}$ is asymptotically vanishing as $h\to-\io$ as $O(\e 2^{\th h})$ from \pref{dec}. It turns out that
\be\tilde\l_{\io}=2(J_3+2\l)(1-\cos 2 p_F)+2\l(1-\cos 4 p_F)+F
\ee
with $|F|\le C\e^2$, and that
$\tilde Z^({0)}
\not =Z^{(1)}$ (the symmetry between space and time is broken in the chain model). In order to understand how \pref{ffbb} is derived we can simply notice that
$K^{(a)11}_\l(\xx)$ is identical to 
$G^{0,2}_{\m,\n}(\pp)$ replacing 
$\l_j, Z_h,Z_{0,h}, Z_{1,h}$ with $\tilde\l_h,\tilde Z_h,\tilde Z_{0,h}, \tilde Z_{1,h}$; to the functions $\hat A_{\m,\n}(\pp)$ contribute terms 
coming from such difference and from 
$K^{(b)11}_\l(\xx)$ in \pref{sacc}. They are all bounded by
\pref{sacc1} summed over the scale and therefore the Fourier transform is {\it continuous} in $\pp$. The r.h.s.  of \pref{ffbb} says that the current-current correlation of the chain
model is equal to the one of a continuous relativistic model, but the contribution
of the irrelevant terms is not small, but simply more regular in Fourier space.  
 
As we noticed in the previous section, from the dimensional bound we cannot exclude
that the Fourier transform current-current correlation of the chain model has 
logarithmic divergences; this can be achieved by using \pref{ffbb}
which allow us to exploit the symmetries of the model \pref{aaaa}.
Indeed while in the spin chain model there is only one set of Ward Identities, 
in the model \pref{aaaa} there are {\it two} set of ward identities, related to the global and chiral symmetry; 
by performing the phase transformations $\psi_{\o, \xx}\to e^{i\a_{\xx}}\psi_{\o,\xx}$ (global phase transformation)
one gets, as proved in \cite{BM}
\bea&&\tilde Z[-i p_0 {1\over \tilde Z^{(0)} } \hat
G^{2,1}_{0;\o}(\kk,\kk+\pp)+  p v_s
{1\over \tilde Z^{(1)}} \hat G^{2,1}_{2;\o}(\kk,\kk+\pp)]=\label{h11}\nn\\
&& = A [\hat G^{2,0}_{\o}(\kk)- \hat G^{2,0}_{\o}(\kk+\pp)]\nn\\
\eea
while performing the phase transformations $\psi_{\o, \xx}\to e^{i\o\a_{\xx}}\psi_{\o,\xx}$ (chiral phase transformation)
it is found
\bea
&&\tilde Z [-i p_0 {1\over \tilde Z^{(1)}} \hat G^{2,1}_{1;\o}(\kk,
\kk+\pp) + v_s
p {1\over \tilde Z^{(0)}} \hat G^{2,1}_{0;\o}(\kk,\kk+\pp)] =\nn\\
&& = \o\bar A [\hat G^{2,0}_{\o}(\kk) - \hat
G^{2,0}_{\o}(\kk+\pp)]\;,\nn \eea
with
\be A^{-1}=1-\t ,\quad \bar A^{-1}=1+\t \quad\quad \t={\tilde\l_\io\over 4\pi v_s} \label{v2a}\ee
Note the presence of $\t$ in the above Ward-Identities, which represent the {\it chiral anomaly}.
In the same way there are two set of Ward Identities for the densities, 
related to the global and chiral transformations;
from them one can write an explicit expressions the density correlations, that is, if
$D_\o(\pp)=-i p_0+\o v_s p$
\bea&&
\hat G^{0,2}_{0,0}= {1\over 4\pi v_s Z^2}{(\tilde
Z^{(0)})^2\over 1-\t^2} [{D_-(\pp)\over
D_+(\pp)}+{D_+(\pp)\over D_-(\pp)}+2 \t ] \label{ggv1}\nn\\
&&\hat G^{0,2}_{1,1}= {1\over 4\pi v_s Z^2}{(\tilde
Z^{(1)})^2\over 1-\t^2} [{D_-(\pp)\over
D_+(\pp)}+{D_+(\pp)\over D_-(\pp)}-2 \t ] \label{w2}
\eea
From the above expression is easy to verify that $G^{0,2}_{\m,\m}(\pp)$
are {\it not} continuous in $\pp$. The density and current correlations are symmetric
between exchange of space and time, contrary to what is expected for the chain model
in which the space and time symmetry is broken by the irrelevant terms. Note also that Ward Identities alone allow
to get an explicit form for the density correlations in the effective model \pref{aaaa}, what is
not possible in the chain model \pref{1a1}.

Eq.\pref{ffbb} provide a relation between the current-current correlation of the chain model \pref{1a1} and of the relativistic model \pref{aaaa}, once that its bare parameters $\tilde\l_\io$, $Z,\tilde Z^{(0)},\tilde Z^{(1)}$
are properly fine tuned. The WI \pref{h11} combined with \pref{cczs}
must coincide with the lattice Ward Identities \pref{ao1};
this provides relations between the bare parameters, namely
\be
{1\over 1-\t}{\tilde Z^{(0)}\over \tilde Z}=1\quad {v_s \tilde Z^{(0)}\over \tilde Z^{(1)}}=1\label{33}
\ee
Let us consider now \pref{ffbb} in which by continuity $\lim_{\pp\to 0} A_{\m\n}(\pp)=A_{\m\n}(0)$ (the limit does not depend from the order contrary to what happens in the first term in the r.h.s. of \pref{ffbb}). The value of
the constant $A_{\m\n}(0)$ is a complicate function depending from all the micro detail of the chain model; however its value is fixed by \pref{ffvv}
\bea
&& 
\hat G^{0,2}_{00}(p_0,0)+\hat A_{00}(p_0,0)=0\nn\\
&&\hat G^{0,2}_{11}(0,p)+\hat A_{11}(0,p)+<j_D>=0
\eea
so that
\bea
&&\hat A_{00}(0,0)=-\lim_{p_0\to 0}\lim_{p\to 0}\hat G^{0,2}_{00}(\pp)\nn\\
&&\hat A_{11}(0,0)+<j_D>=-\lim_{p\to 0}\lim_{p_0\to 0}\hat G^{0,2}_{11}(\pp)\label{34}
\eea
Note that $\hat A_{00}(0,0)$ and $A_{11}(0,0)$, containing the contributions from the irrelevant terms, are essential for the result; they break the symmetry between space and time which is present in the effective model \pref{aaaa} which
is not true in the chain model \pref{1a1}; as a result the limits $\lim_{p\to 0}$
and $\lim_{p_0\to 0}$ do not commute in the current correlations.

Therefore by 
\pref{ffbb}, \pref{w2} ,\pref{34}
\bea
&&\hat K^{00}_\l(\pp) ={1\over \pi v_s Z^2}{(\tilde
Z^{(0)})^2\over 1-\t^2}{v_s^2 p^2\over
p_0^2+v_s^2 p^2}+O(\pp)\\
&&\hat D_\l(\pp)={1\over \pi v_s Z^2}{(\tilde
Z^{(1)})^2\over 1-\t^2}{p_0^2\over
p_0^2+v_s^2 p^2}+O(\pp)
\eea
and using \pref{33}
\bea
&&\hat K^{00}_\l(\pp) ={K\over \pi v_s}{v_s^2 p^2\over
p_0^2+v_s^2 p^2}+O(\pp)\\
&&\hat D_\l(\pp)={K v_s\over \pi}{p_0^2\over
p_0^2+v_s^2 p^2}+O(\pp)
\eea
with $K={1-\t\over 1+\t}$, and using that $\k_\l=\lim_{p\to 0}\lim_{p_0\to 0}
\hat K^{00}_\l(\pp) $ and $D_\l=\lim_{p_0\to 0}\lim_{p\to 0}
\hat D_\l(\pp) $ finally \pref{aa1} follows.

\section{Conclusions}

We have rigorously established a strict lower bound for the zero temperature Drude weight which is strictly positive for the 
the $XXZ$ chain perturbed by a weak next to nearest neighbor interaction \pref{1a1}, proving that
the system has an ideal metallic behavior (infinite conductivity). We have also proved that the Drude weight verifies the same exact relation \pref{aass} as in absence
of the integrability breaking perturbation. The main difficulty in the analysis relies in the irrelevant terms, whose role is essential in ensuring the correct lattice symmetries, and this makes the use of the ERG quite well suited for the problem. The results are obtained exploiting both the lattice and the emerging chiral symmetries of the theory, and relying on rigorous estimates
on the large distance decay of the correlations, based on determinant bounds.
Our results are conclusive regarding the properties of the Drude weight in
a non integrable spin chain at {\it zero} temperature, at least for weak perturbations, 
and the main problem remains the conductivity properties at  {\it non zero} temperature. 
The main difficulty of such a problem relies on the subtle interplay between dangerously
irrelevant terms and conserved quantities; therefore we believe that the understanding of such an issue at zero temperature, which is achieved in the present paper,
is an essential prerequisite for an analytical understanding for the positive temperature problem.

\end{document}